# Transaction-Oriented Simulation in Ad Hoc Grids

Gerald Krafft
Harrow School of Computer Science, University of Westminster
Watford Rd, Northwick Park, Harrow HA1 3TP, U.K.
g.krafft@gmx.net

**Abstract**

*This paper analyses the requirements of performing parallel transaction-oriented simulations with a special focus on the space-parallel approach and discrete event simulation synchronisation algorithms that are suitable for transaction-oriented simulation and the target environment of Ad Hoc Grids. To demonstrate the findings a Java-based parallel transaction-oriented simulator for the simulation language GPSS/H is implemented on the basis of the most promising Shock Resistant Time Warp synchronisation algorithm and using the Grid framework ProActive. The validation of this parallel simulator shows that the Shock Resistant Time Warp algorithm can successfully reduce the number of rolled back Transaction moves but it also reveals circumstances in which the Shock Resistant Time Warp algorithm can be outperformed by the normal Time Warp algorithm. The conclusion of this paper suggests possible improvements to the Shock Resistant Time Warp algorithm to avoid such problems.*

## 1. Introduction

The growing demand of complex Computer Simulations for instance in engineering, military, biology and climate research has also lead to a growing demand in computing power. One possibility to reduce the runtime of large, complex Computer Simulations is to perform such simulations distributed on several CPUs or computing nodes. This has induced the availability of high-performance parallel computer systems. Even so the performance of such systems has constantly increased, the ever-growing demand to simulate more and more complex systems means that suitable high-performance systems are still very expensive.

Grid computing promises to provide large-scale computing resources at lower costs by allowing several organisations to share their resources. But traditional Computing Grids are relatively static environments that require a dedicated administrative authority and are therefore less well suited for transient short-term collaborations and small organisations with fewer resources. Ad Hoc Grids provide such a dynamic and transient resource-sharing infrastructure [8] that allows even small organisations or individual users to form Computing Grids. They will make Grid computing and Grid resources widely available to small organisations and mainstream users allowing them to perform resource intensive computing tasks like Computer Simulations.

There are several approaches to performing Computer Simulations distributed across a parallel computer system. The space-parallel approach [3] is one of these approaches that is robust, applicable to many different simulation types and that can be used to speed up single simulation runs. It requires the simulation model to be partitioned into relatively independent sub-systems that are then performed in parallel on several nodes. Synchronisation between these nodes is still required because the model sub-systems are usually not fully independent. A lot of past research has concentrated on different synchronisation algorithms for parallel simulation. Some of these are only suitable for certain types of parallel systems, like for instance shared memory systems.

This work investigates the possibility of performing parallel transaction-oriented simulation in an Ad Hoc Grid environment with the main focus on the aspects of parallel simulation. Potential synchronisation algorithms and other simulation aspects are analysed in respect of their suitability for transaction-oriented simulation and Ad Hoc Grids as the target environment and the chosen solutions are described and reasons for their choice given. To demonstrate the solutions a Java-based parallel transaction-oriented simulator for the simulation language GPSS/H [7] is implemented and evaluated using a set of simple example simulation models.

## 2. Synchronisation Algorithm

The choice of synchronisation algorithm can have a significant influence on how much of the parallelism that exists in a simulation model will be utilised by the parallel simulation system. Conservative algorithms utilise the parallelism less well than optimistic algorithms because they require guarantees, which in most cases are derived from additional knowledge about the behaviour of the simulation model, like for instance the communication topology or lookahead attributes of the model. For this reason conservative algorithms are often used to simulate very specific systems where such knowledge is given or can easily be derived from the model. For general simulation systems optimistic algorithms are better suited as they can utilise the parallelism within a model to a higher degree without requiring any guarantees or additional knowledge.

Another important aspect of choosing the right synchronisation algorithm is the relation between the performance properties of the expected parallel hardware architecture and the granularity of the parallel algorithm. In order for the parallel algorithm to perform well in general on the target hardware environment the granularity of the algorithm, i.e. the ratio between computation and communication has to fit the ratio of the computation performance and communication performance provided by the parallel hardware.

### 2.1. Requirements

Considering the target environment of Ad Hoc Grids and the goal of designing and implementing a general parallel simulation system based on the transaction-oriented simulation language GPSS/H it can be concluded that the best suitable synchronisation algorithm is an optimistic or hybrid algorithm that has a coarse grained granularity. The algorithm should require only little communication compared to the amount of computation it performs. At the same time the algorithm needs to be flexible enough to adapt to a changing environment, as this is the case in Ad Hoc Grids. A further requirement is that the algorithm can be adapted to and is suitable for transaction-oriented simulation.

### 2.2. Chosen Algorithm

The most promising algorithm found for these requirements is the Shock Resistant Time Warp algorithm [2]. This algorithm has some similarities with the Elastic Time algorithm [9] and also the Adaptive Memory Management algorithm [1] but at the same time is suitable for loosely coupled parallel systems like Computing Grids. Similar to the Elastic Time algorithm state vectors are used to describe the current states of all LPs plus a set of functions to determine the output vector but the Shock Resistant Time Warp algorithm does not require a global state. Instead each LP separately tries to optimise its parameters towards the best performance. And similar to the Adaptive Memory Management algorithm the optimism is controlled indirectly be setting artificial memory limits but for the Shock Resistant Time Warp algorithm each LP will limit its own memory instead of using an overall memory limit for the whole simulator.

The Shock Resistant Time Warp algorithm is a fully distributed approach to controlling the optimism in Time Warp that requires no additional communication between the LPs [2]. It is based on the Time Warp algorithm [6] but extends each LP with a control component called LPCC that constantly collects information about the current state of the LP using a set of sensors. These sets of sensor values are then translated into sets of indicator values representing state vectors for the LP. The LPCC will keep a history of such state vectors using a clustering technique so that it can search for past state vectors that are similar to the current state vector but provide a better performance indicator. An actuator value will be derived from the most similar of such state vectors that is subsequently used to control the optimism of the LP.

As the Shock Resistant Time Warp algorithm was designed for discrete event simulation its sensors and indicators had to be adapted to the equivalent values in transaction-oriented simulation.

## 3. End of Simulation

Another important aspect that had to be considered is the detection and correct handling of the simulation end. In transaction-oriented simulation a simulation is complete when the defined end state is reached, i.e. the termination counter reaches a value less or equal to zero. When using an optimistic synchronisation algorithm for the parallelisation of transaction-oriented simulation it is crucial to consider that optimistic algorithms will first execute all local events without guarantee that the causal order is correct. They will recover from wrong states by performing a rollback if it later turns out that the causal order was violated. Therefore any local state reached by an optimistic LP has to be considered provisional until a GVT has been received that guarantees the state. In addition it is necessary to bear in mind that at any point in real time it is most likely that each of the LPs has reached a different local simulation time so that after an end state has been reached by one of the LPs that is guaranteed by a GVT it is important to synchronise the states of all LPs so that the combined end state from all model partitions is equivalent to the model end state that would have been reached in a sequential simulator.

The mechanism suggested here leads to a consistent and correct global end state of the simulation considering the problems mentioned above. For this mechanism the LP reaching a provisional end state is switched into the provisional end mode. In this mode the LP will stop to process any further Transactions leaving the local model partition in the same state but it will still respond to and process control messages like GVT parameter requests and it will receive Transactions from other LPs that might cause a rollback. The LP will stay in this provisional end mode until the end of the simulation is confirmed by a GVT or a received Transaction causes a rollback with a potential re-execution that is not resulting in the same end state. While the LP is in the provisional end mode additional GVT parameters are passed on for every GVT calculation denoting the fact that a provisional end state has been reached and the simulation time and priority of the Transaction that caused the provisional end. The GVT calculation process can then assess whether the earliest current provisional end state is guaranteed by the GVT. If this is the case then all other LPs are forced to synchronise to the correct end state by rolling back using the simulation time and priority of the Transaction that caused the provisional end and the simulation is stopped.

## 4. Suitable Cancellation Technique

Transaction-oriented simulation has some specific properties compared to discrete event simulation. One of these properties is that Transactions do not consume simulation time while they are moving from block to block. This has an influence on which of the synchronisation algorithms are suitable for transaction-oriented simulation but also on the cancellation techniques used. If a Transaction moves from LP1 to LP2 then it will arrive at LP2 with the same simulation time that it had at LP1. A Transaction moving from one LP to another is therefore equivalent to an event in discrete event simulation that when executed creates another event for the other LP with exactly the same time stamp. Because simulation models can contain loops, as it is for instance common for models of quality control systems where an item failing the quality control needs to loop back through the production process, this specific behaviour of transaction-oriented simulation can lead to endless rollback loops for certain cancellation techniques.

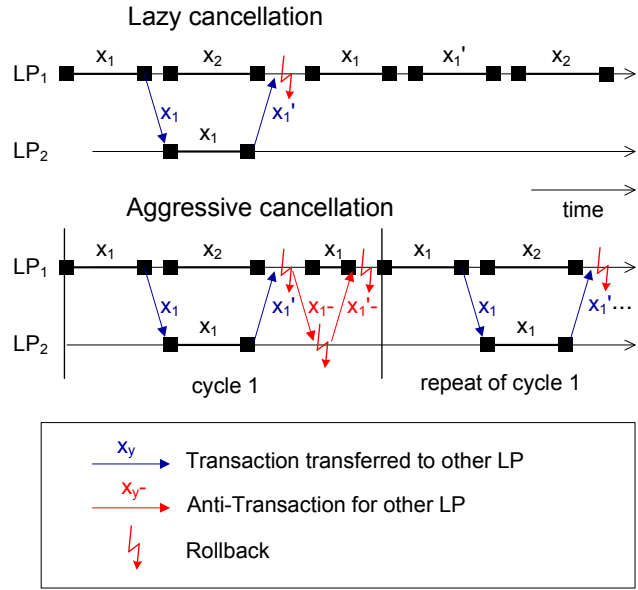

Figure 1: Cancellation in transaction-oriented simulation

Figure 1 compares the rollback behaviour of aggressive cancellation and lazy cancellation [4] in respect of such a loop within the simulation model. It shows the movement of Transaction $x_1$ from $LP_1$ to $LP_2$ but without a delay in simulation time the Transaction is transferred back to $LP_1$. As a result $LP_1$ will be rolled back to the simulation time just before $x_1$ was moved. At this point two copies of Transaction $x_1$ will exist in $LP_1$. The first one is $x_1$ itself which needs to be moved again and the second is $x_1$' which is the copy that was send back from $LP_2$. This is the point from where the execution differs between lazy cancellation and aggressive cancellation. In lazy cancellation $x_1$ would be moved again resulting in the same transfer to $LP_2$. But because $x_1$ was sent to $LP_1$ already it will not be transferred again and no anti-transaction will be sent. From here $LP_1$ just proceeds moving the Transactions in its Transaction chain according to their simulation time (Transaction priorities are ignored for this example). Apposed to that the rollback in aggressive cancellation would result in an anti-Transaction being sent out for $x_1$ immediately which would cause a second rollback in $LP_2$ and another anti-Transaction for $x_1$' being sent back to $LP_1$. At the end both LPs will end up in the same state in which they were before $x_1$ was moved by $LP_1$. The same cycle of events would start again without any actual simulation progress.

In order to avoid the described endless rollback loops lazy cancellation needs to be used for parallel transaction-oriented simulation.

## 5. Implementation

The parallel transaction-oriented simulator was implemented using the Java-based Grid environment ProActive [5] that is very well suited for Ad Hoc Grids. The overall architecture of the parallel simulator follows the Master-Slave approach. Figure 2 shows the simplified architecture of the parallel simulator including its main components.

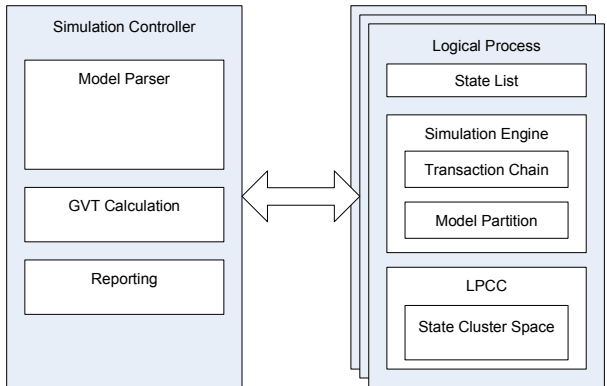

Figure 2: Architecture overview

The main parts of the parallel simulator are the Simulation Controller and the Logical Processes. The Simulation Controller controls the overall simulation. It is created when the user starts the simulation and will use the Model Parser component to read the simulation model file and parse it into an in memory object structure representation of the model. After the model is parsed the Simulation Controller creates Logical Process instances, one for each model partition contained within the simulation model. The Simulation Controller and the Logical Processes are implemented as ProActive Active Objects so that they can communicate with each other via method calls. Communication will take place between the Simulation Controller and the Logical Processes but also between the Logical Processes for instance in order to exchange Transactions. Note that the communication between the Logical Processes is not illustrated in Figure 2. After the Logical Process instances have been created and initialised they will receive the model partitions that they are going to simulate from the Simulation Controller and the simulation is started. Each Logical Process implements an LP according to the Shock Resistant Time Warp algorithm. The main component of the Logical Process is the Simulation Engine, which contains the Transaction chain and the model partition that is simulated. The Simulation Engine is the part that is performing the actual simulation. It is moving the Transactions from block to block by executing the block functionality using the Transactions. Another important part of the Logical Process is the State List. It contains historic simulation states in order to allow rollbacks as required by optimistic synchronisation algorithms. Note that Figure 2 does not show other lists within the Simulation Engine like for instance the list of Transactions received and the list of Transactions sent to other Logical Processes. Furthermore the Logical Process will contain the Logical Process Control Component (LPCC) that is part of the Shock Resistant Time Warp algorithm. An option to disable the LPCC will allow for the simulator to either operate in Shock Resistant Time Warp mode or in normal Time Warp mode.

The Simulation Controller will perform GVT calculations in order to establish the overall progress of the simulation and in order to reclaim memory through fossil collection if requested by one of the Logical Processes. GVT calculation will also be used to confirm a provisional simulation end state that might be reached by one of the Logical Processes.

When the end of the simulation is reached then the Simulation Controller will ensure that the partial models in all Logical Processes are set to the correct and consistent end state and it will collect information from all Logical Processes in order to assemble and output the post simulation report.

## 6. Simulation results

A set of example simulation models was used to validate the parallel simulator and to compare the Shock Resistant Time Warp algorithm with the normal Time Warp algorithm. The simulation models were deliberately kept very simple in order to evaluate conceptual aspects of the parallel simulator. The validation runs were performed on a standard PC with a single CPU (Intel Pentium 4 with 3.2GHz, 1GB RAM) running SuSE Linux 10.0. Even so the simulation runs could only be performed on a single CPU, the results allow some significant performance conclusions to be drawn for the parallel simulator and the employed synchronisation algorithm.

### 6.1. Reduction of Rolled Back Transaction Moves

The simulation model used for this evaluation contains two partitions. Both partitions have a GENERATE block and a TERMINATE block but in addition partition 1 also contains a TRANSFER block that with a very small probability of 0.001 sends some of its Transactions to partition 2. The whole model is constructed so that partition 2 is usually ahead in simulation time compared to partition 1, achieved through the different configuration of the GENERATE blocks, and that occasionally partition 2 receives a Transaction from the first partition. Because partition 2 is usually ahead in simulation time this will lead to rollbacks in this partition. The simulation stops when 20000 Transactions have been terminated in partition 2. This model attempts to emulate the common scenario where a distributed simulation uses nodes with different performance parameters or partitions that create different loads so that during the simulation the LPs drift apart and some of them are further ahead in simulation time than others leading to rollbacks and re-execution. The details of the model used can be seen below.

```
PARTITION Partition1,20000
GENERATE 1,0
TRANSFER 0.001,Label1
TERMINATE 0
PARTITION Partition2,20000
GENERATE 4,0,5000
Label1 TERMINATE 1
```

This model was simulated once in Shock Resistant Time Warp mode and once in normal Time Warp mode by enabling or disabling the LPCC within the simulator. The output of the simulation runs show that in Shock Resistant Time Warp mode the LPCC successfully reduced the number of rolled back transaction moves compared to the normal Time Warp mode by limiting the number of uncommitted transaction moves using the actuator. The graph below indicates how the LPCC adapts the actuator value during the simulation.

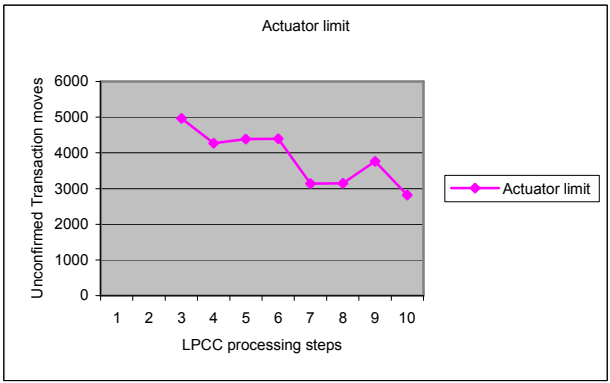

Figure 3: Actuator value graph

And the following table shows the actual reduction in the number of rolled back transaction moves within LP2.

| LP statistic item | LPCC on | LPCC off |
|---|---|---|
| Total committed Transaction moves | 19639 | 19953 |
| Total Transaction moves rolled back | 70331 | 77726 |
| Total simulated Transaction moves | 90330 | 97725 |

Table 1: LP2 processing statistics

Table 1 shows that the simulation run using the Shock Resistant Time Warp algorithm (LPCC on) required around 7400 less rolled back Transaction moves, which is about 10% less compared to the simulation run using the normal Time Warp algorithm (LPCC off). As a result the total number of Transaction moves performed by the simulation was reduced as well.

## 6.1. Time Warp outperforming Shock Resistant Time Warp

During the testing of the parallel simulator it became apparent that in same cases the normal Time Warp algorithm can outperform the Shock Resistant Time Warp algorithm. This second evaluation demonstrates this in an example. The simulation model used is very similar to the one used for the last evaluation. It contains two partitions with the first partition transferring some of its Transactions to the second partition but this time the GENERATE blocks are configured so that the first partition is ahead in simulation time compared to the second. The simulation is finished when 3000 Transactions have been terminated in one of the partitions. The complete simulation model can be seen here:

```
PARTITION Partition1,3000
GENERATE 1,0,2000
TRANSFER 0.3,Label1
TERMINATE 1
PARTITION Partition2,3000
GENERATE 1,0
Label1 TERMINATE 1
```

As a result of the changed GENERATE block configuration and the first partition being ahead of the second partition in simulation time, all Transactions received by partition 2 from partition 1 are in the future for partition 2 and no rollbacks will be caused. But it will lead to an increase of the number of outstanding Transactions within partition 2 pushing up the number of uncommitted Transaction moves during the simulation.

The first simulation run was performed with the LPCC on, i.e. in Shock Resistant Time Warp mode. The significant effect of the simulation run is that the LPCC in LP2 starts setting actuator values in order to steer the local simulation processing towards a state that promises better performance but because the number of uncommitted Transaction moves within the second partition increases as a result of the Transactions received from partition 1 the actuator limits set by the LPCC are reached and the LP is switched into cancelback mode leading to its simulation progress being slowed down. In addition LP1 is also slowed down by the Transactions cancelled back from LP2. LP2 keeps switching into cancel back mode and keeps cancelling back Transactions to LP1 for large parts of the simulation resulting in a significant slowdown of the overall simulation progress. As a result the simulation run that took 16.5s in normal Time Warp mode took 27.3s in Shock Resistant Time Warp mode.

## 7. Conclusion

This work briefly discussed the requirements for a synchronisation algorithm suitable for Ad Hoc Grid environments as well as transaction-oriented simulation. Further requirements for parallel transaction-oriented simulation were analysed and possible solutions suggested. The Shock Resistant Time Warp algorithm was chosen as the most promising algorithm that fulfils the requirements. The algorithm was adapted to transaction-oriented simulation and a parallel simulator was implemented using the Grid environment ProActive. The parallel simulator can operate in Shock Resistant Time Warp mode as well as normal Time Warp mode allowing comparison of the two algorithms for different transaction-oriented simulation models.

The evaluation of the parallel simulator showed that the Shock Resistant Time Warp algorithm can successfully reduce the number of rolled back Transaction moves, which for simulations with many or long cascaded rollbacks will lead to a better simulation performance. But it also revealed a weakness of the Shock Resistant Time Warp algorithm. Because LPs try to optimise their properties based only on local information it is possible for the Shock Resistant Time Warp algorithm to perform significantly worse than the normal Time Warp algorithm. Future work on this simulator could improve the Shock Resistant Time Warp algorithm by making the LPs aware of their position within the global progress of the simulation.